%% file: Main.tex
\documentclass[10pt,twocolumn,letterpaper]{article}

\usepackage{graphicx}
\usepackage{amsmath}
\usepackage{amssymb}
\usepackage{booktabs}
\usepackage{times}
\def\X{{\mathbf X}}

\usepackage[pagebackref,breaklinks,colorlinks]{hyperref}

\usepackage[capitalize]{cleveref}
\crefname{section}{Sec.}{Secs.}
\Crefname{section}{Section}{Sections}
\Crefname{table}{Table}{Tables}
\crefname{table}{Tab.}{Tabs.}

\def\cvprPaperID{6193} 
\def\confName{CVPR}
\def\confYear{2022}

\begin{document}

\title{Seeing through a Black Box: Toward High-Quality Terahertz Tomographic Imaging via Multi-Scale Spatio-Spectral Image Fusion}
\author{{Weng-tai Su$^{\dag}$, Yi-Chun Hung$^{\dag}$, Ta-Hsuan Chao$^{\dag}$, Po-Jen Yu$^{\dag}$, Shang-Hua Yang$^{\dag}$ and Chia-Wen Lin$^{\dag}$}\\
$^{\dag}$National Tsing Hua University, Taiwan
}

\maketitle

\begin{abstract}
Terahertz (THz) imaging has recently attracted significant attention thanks to its non-invasive, non-destructive, non-ionizing, material-classification, and ultra-fast nature for object exploration and inspection. However, its strong water absorption nature and low noise tolerance lead to undesired blurs and distortions of reconstructed THz images. The performances of existing restoration methods are highly constrained by the diffraction-limited THz signals.  To address the problem, we propose a novel Subspace-and-Attention-guided Restoration Network (\texttt{SARNet}) that  fuses  multi-spectral features of a THz image for effective restoration. To this end, \texttt{SARNet} uses multi-scale branches to extract spatio-spectral features of amplitude and phase which are then  fused via shared subspace projection and attention guidance. Here, we experimentally construct ultra-fast THz time-domain spectroscopy system  covering a broad frequency range from 0.1 THz to 4 THz for building up temporal/spectral/spatial/phase/material  THz  database  of hidden 3D objects. Complementary to a quantitative evaluation, we demonstrate the effectiveness of our \texttt{SARNet} model on 3D THz tomographic reconstruction applications.
\end{abstract}
\vspace{-0.3in}

\section{Introduction}
\label{sec:intro}

\input{intro}

\section{Related Work}
\label{sec:related}

\input{related}

\section{Terahertz Image Restoration}
\label{sec:method}
\input{method}
\section{Experiments}
\label{sec:experiments}
\input{experiments}

\section{Conclusion}
\label{sec:conclusion}
\input{conclusion}

{\small
\bibliographystyle{ieee_fullname}
\bibliography{terahertz}
}

\end{document}

%% file: intro.tex
Ever since the first camera's invention, imaging under different bands of electromagnetic (EM) waves, especially X-ray and visible lights, has revolutionized our daily lives \cite{kamruzzaman2011application, rotermund1991methods, yujiri2003passive}. X-ray imaging plays a crucial role in medical diagnosis, such as cancer, odontopathy, and COVID-19 symptom \cite{abbas2021classification, round2005preliminary, tuan2018dental}, based on X-ray's high penetration depth to great varieties of materials; visible-light imaging has not only changed the way of recording lives but contributes to the development of artificial intelligence (AI) applications, such as surveillance security and surface defect inspection~\cite{xie2008review}. However, X-ray and visible-light imaging still face tough challenges. X-ray imaging is ionizing, which is harmful to biological objects and thus severely limits its application scope \cite{de2004risk}. On the other hand, although both non-ionizing and non-destructive, visible-light imaging cannot retrieve most optically opaque objects' interior information due to the highly absorptive and intense scattering behaviors between light and matter in the visible light band. To visualize the 3D information of objects in a remote but accurate manner, terahertz (THz) imaging has become among the most promising candidates among all EM wave-based imaging techniques
~\cite{abraham2010non, yu2012potential}.

\begin{figure}[t]
\centering
\includegraphics[width=0.5\textwidth]{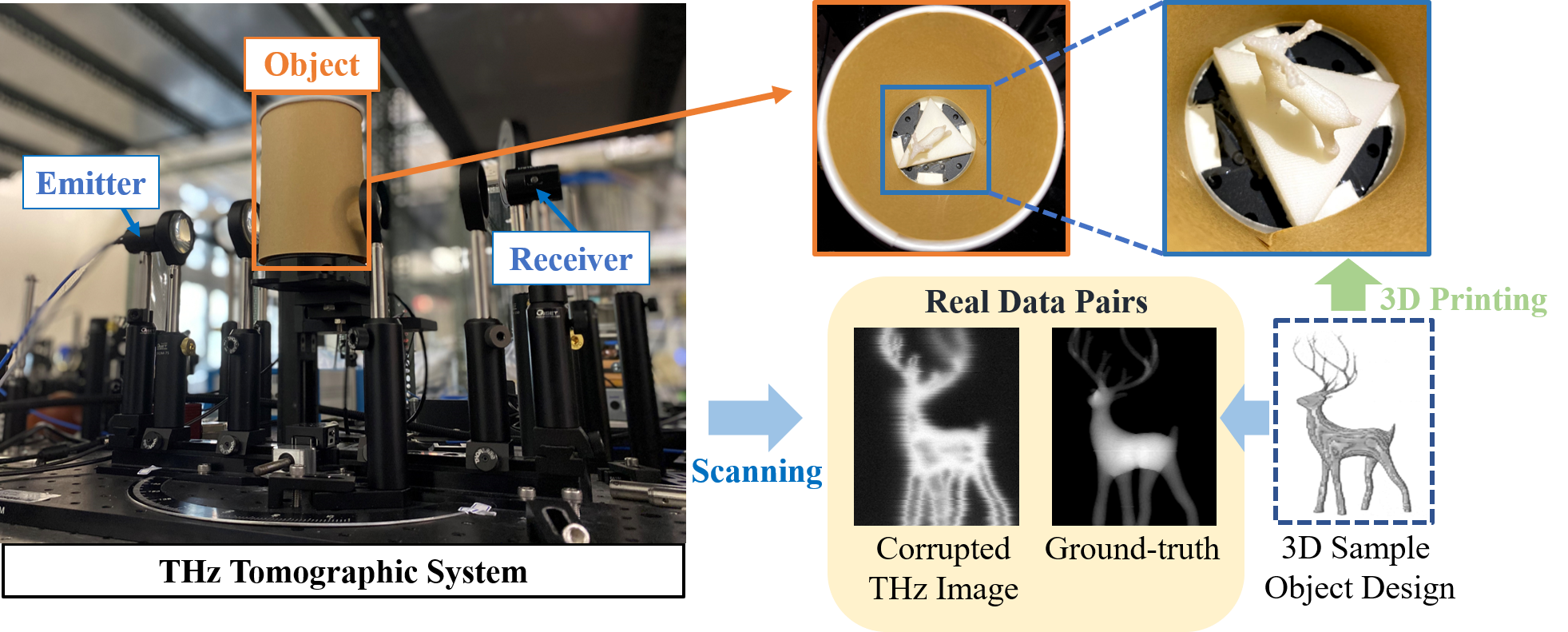}
\vspace{-0.3in}
\caption{Illustration of THz data collection with our in-house THz-TDS tomographic imaging system.} 
\label{fig:design}	
\vspace{-0.3 in}
\end{figure}


\begin{figure*}[t]
\centering
\includegraphics[width=0.68\textwidth]{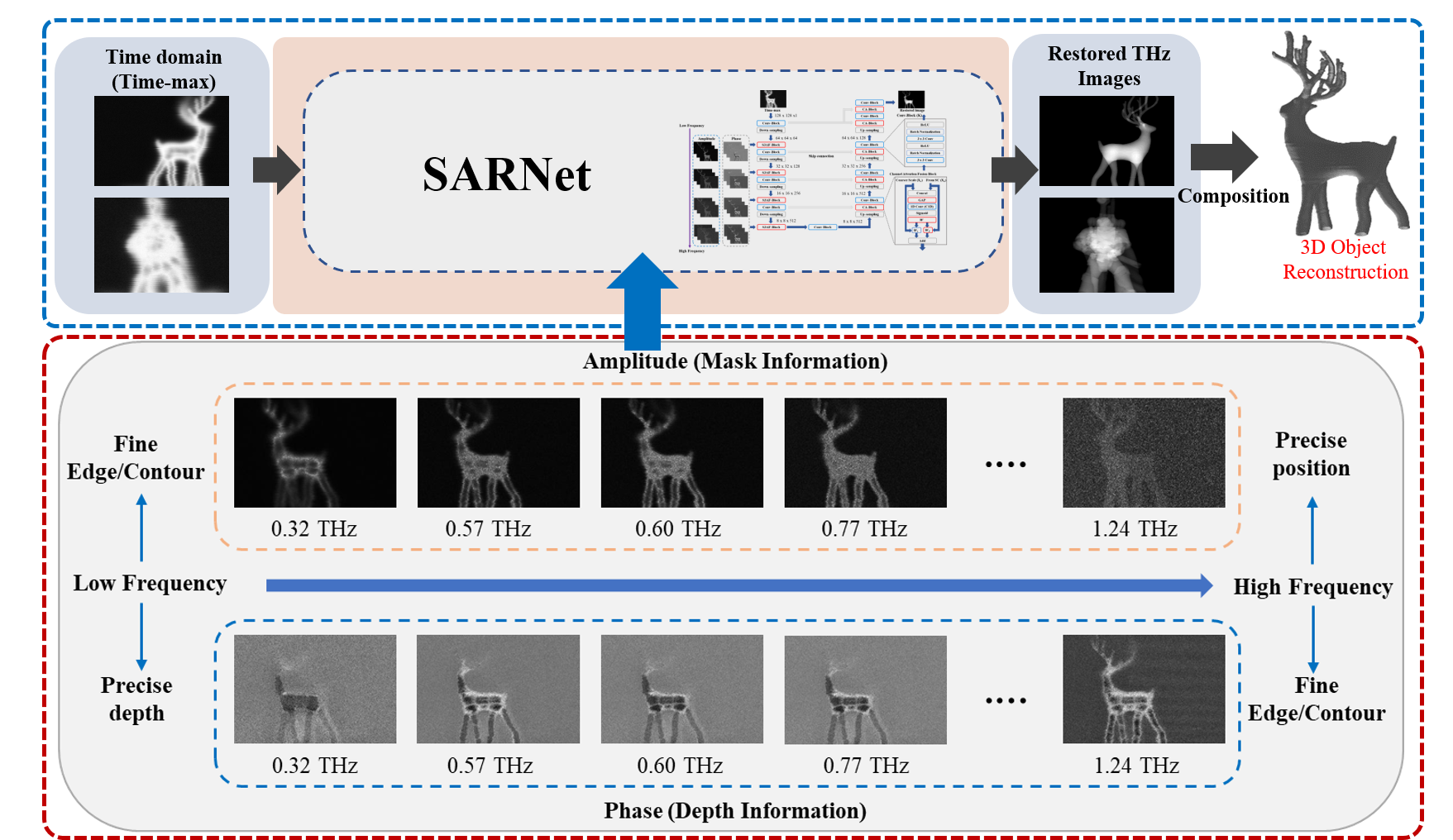}
\caption{Illustration of THz 3D tomographic imaging based on \texttt{SARNet}. The red block shows the images retrieved from different spectral frequencies  of the amplitude and phase of the THz measurements of \textbf{Deer}.} 
\label{fig:band}	
\vspace{-0.2in}
\end{figure*}

THz radiation, between microwave and infrared, has often been regarded as the last frontier of EM wave~\cite{saeedkia2013handbook}, which provides its unique functionalities among all EM bands. Along with the rapid development of THz technology, THz imaging has recently attracted significant attention due to its non-invasive, non-destructive, non-ionizing, material-classification, and ultra-fast nature for advanced material exploration and engineering. As THz waves can partially penetrate through varieties of optically opaque materials, it carries hidden material tomographic information along the traveling path, making this approach a desired way to see through black boxes without damaging the exterior \cite{mittleman1999recent, jansen2010terahertz, mittleman2018twenty}. By utilizing light-matter interaction within the THz band, multifunctional tomographic information of a great variety of materials can also be retrieved even at a remote distance. In the past decades, THz time-domain spectroscopy (THz-TDS) has become one of the most representative THz imaging modalities to achieve non-invasive evaluation because of its unique capability of extracting geometric and multi-functional information of objects. Owing to its fruitful information in multi-dimensional domains --- space, time, frequency, and phase, THz-TDS imaging has been already allocated for numerous emerging fields, including drug detection~\cite{kawase2003non}, industrial inspection, cultural heritage inspection~\cite{fukunaga2016thz} and cancer detection~\cite{bowman2018pulsed}.


To retrieve temporal-spatio-spectral information of each object voxel, our THz imaging experiment setup is based on a THz-TDS system as shown in Fig.~\ref{fig:design}. Our measured object (a covered 3D printed deer) is placed on the rotation stage in the THz path between the THz source and detector of the THz-TDS system. During the scanning, the THz-TDS system profiles each voxel's THz temporal signal with 0.1 ps temporal resolution, whose amplitude corresponds to the strength of THz electric field. Based on the dependency between the amplitude of a temporal signal and THz electric field, in conventional THz imaging, the maximum peak of the signal ($\texttt{Time-max}$) is extracted as the feature for a voxel. The reconstructed image based on $\texttt{Time-max}$ features can deliver great signal-to-noise ratio and a clear object contour. However, the conventional THz imaging based on $\texttt{Time-max}$ features suffers from several drawbacks, such as the undesired contour in the boundary region, the hollow in the body region, and the blurs in high spatial-frequency regions. To break this limitation, we utilize the spectral information of THz temporal signals to supplement the conventional method based on $\texttt{Time-max}$ features since the voxel of the material behaviors are encoded in both the phase and amplitude of different frequency components, according to the Fresnel equation \cite{dorney2001material}.

Due to the large number of spectral bands with measured THz image data, it is required to sample a subset of the spectral bands to reduce the training parameters. The THz beam is significantly attenuated at water absorption frequencies. Thus, reconstructed THz images based on water absorption lines offer worse details. Besides, the high-speed THz-TDS system offers more than 20 dB SNR in a frequency range of 0.3 THz--1.3 THz. Considering the water absorption in THz regime \cite{van1989terahertz, slocum2013atmospheric} and the superior SNR in the range of 0.3 THz--1.3 THz, we select 12 frequencies at 0.380, 0.448, 0.557, 0.621, 0.916, 0.970, 0.988, 1.097, 1.113, 1.163, 1.208, and 1.229 THz.
The spectral information including both amplitude and phase at the selected frequencies is extracted and then employed to restore clear 2D images. Specifically, Fig.~\ref{fig:band} shows multiple 2D THz images of the same object at the selected frequencies, showing very different contrasts and spatial resolutions as these hyperspectral THz image sets have different physical characteristics through the interaction of THz waves with objects. 
The lower-frequency phase images offer relatively accurate depth information due to their higher SNR level, whereas the higher-frequency phase images offer finer contours and edges because of the shrinking diffraction-limited wavelength sizes (from left to right in Fig.~\ref{fig:band}). The phase also contains, however, a great variety of information of light-matter interaction that could cause learning difficulty for the image restoration task. 
%
To address this issue, we utilize amplitude spectrum as complementary information. Although the attenuated amplitude spectrum cannot reflect comparable depth accuracy levels as phase spectrum, amplitude spectrum still present superior SNR and more faithful contours such as the location information of a measured object. Besides, as the complementary information to phase, the lower-frequency amplitude offers higher contrast, whereas the higher-frequency amplitude offers a better object mask.

In summary, the amplitude complements the shortcomings of phase. The advantages of fusing the two signals from low-frequency to high-frequency are as follows: Since the low-frequency THz signal provides precise depth (the thickness of an object) and fine edge/contour information in the phase and amplitude, respectively, they together better delineate and restore the object. In contrast, the high-frequency feature maps of amplitude and phase respectively provide better edges/contours and precise position information, thereby constituting a better object mask from the complementary features. With these multi-spectral properties of THz images, we can extract rich information from a wide spectral range in the frequency domain to simultaneously restore the 2D THz images without any additional computational cost or equipment, which is beneficial for further development of THz imaging.

We here propose a multi-scale \textbf{S}ubspace-and-\textbf{A}ttention guided \textbf{R}estoration \textbf{Net}work (\texttt{SARNet}) that fuses complementary spectral features of the THz amplitude and phase to supplement the \texttt{Time-max} image for restoring clear 2D images. To this end, \texttt{SARNet} learns common representations in a common latent subspace shared between the amplitude and phase, and then incorporates a Self-Attention mechanism to learn the short/long-range dependency of the spectral features for guiding the restoration task.  Finally,  from clear 2D images restored from corrupted images of an object captured from different angles, we can reconstruct high-quality 3D tomography via inverse Radon transform. Our main contributions are summarized as follows:


\begin{itemize}
\item We are the first research group to merge THz temporal-spatio-spectral data, data-driven models, and material properties to the best of our knowledge. The proposed \texttt{SARNet} has an excellent performance in extracting and fusing features from the light-matter interaction data in THz spectral regime, which inherently contains fruitful 3D object information and its material behaviors. Based on the delicately designed architecture of \texttt{SARNet} on feature fusion, \texttt{SARNet} delivers state-of-the-art performance on THz image restoration.



\item \noindent With our newly established THz tomography dataset --- the world's first in its kind, we provide comprehensive quantitative/qualitative analyses among \texttt{SARNet} and SOTAs. \texttt{SARNet} significantly outperforms  \texttt{Time-max},  baseline  U-Net,  and  multi-spectral  U-Net  by   11.17dB,  2.86dB, and 1.51dB in average PSNR.

\item  This pioneering work shows that computer vision techniques can significantly contribute to the THz community and further open up a new interdisciplinary research field to boost practical applications, e.g., non-invasive evaluation, gas tomography,  industrial inspection, material exploration, and biomedical imaging.
\end{itemize}


%% file: related.tex


\subsection{Deep Learning-based Image Restoration}
In recent years, deep learning methods were first popularized in high-level visual tasks, and then gradually penetrated into many tasks such as image restoration and segmentation. 
Convolutional neural network (CNNs) have  proven to achieve the state-of-the-art performances in fundamental image restoration problem \cite{zhang2017beyond, mao2016image, zhang2020residual, zhang2018ffdnet, ronneberger2015u}. Several network models for image restoration were proposed, such as U-Net \cite{ronneberger2015u}, hierarchical residual network \cite{mao2016image} and residual dense network \cite{zhang2020residual}. Notably, DnCNN \cite{zhang2017beyond} uses convolutions, BN, and ReLU to build 17-layer network for image restoration which was not only utilized for blind image denoising, but was also employed for image super-resolution and JPEG image deblocking. FFDNet \cite{zhang2018ffdnet} employs noise level maps as inputs and utilizes a single model to develop variants for solving problems with multiple noise levels. In \cite{mao2016image} a very deep residual encoding-decoding (RED) architecture was proposed  to solve the image restoration problem using skip connections. \cite{zhang2020residual} proposed a residual dense network (RDN), which maximizes the reusability of features by using residual learning and dense connections. NBNet \cite{cheng2021nbnet} employs subspace projection to transform learnable feature maps into the projection basis, and leverages non-local image information to restore local image details.
Similarly, the \texttt{Time-max} image obtained from a THz imaging system can be cast as an image-domain learning problem which was rarely studied due to the difficulties in THz image data collection. Research works on image-based THz imaging include \cite{popescu2010point, popescu2009phantom, wong2019computational}, and THz tomographic imaging works include \cite{hung2019terahertz, hung2019kernel}.  

\begin{figure*}[!hbt]
\centering
\includegraphics[width=0.92\textwidth]{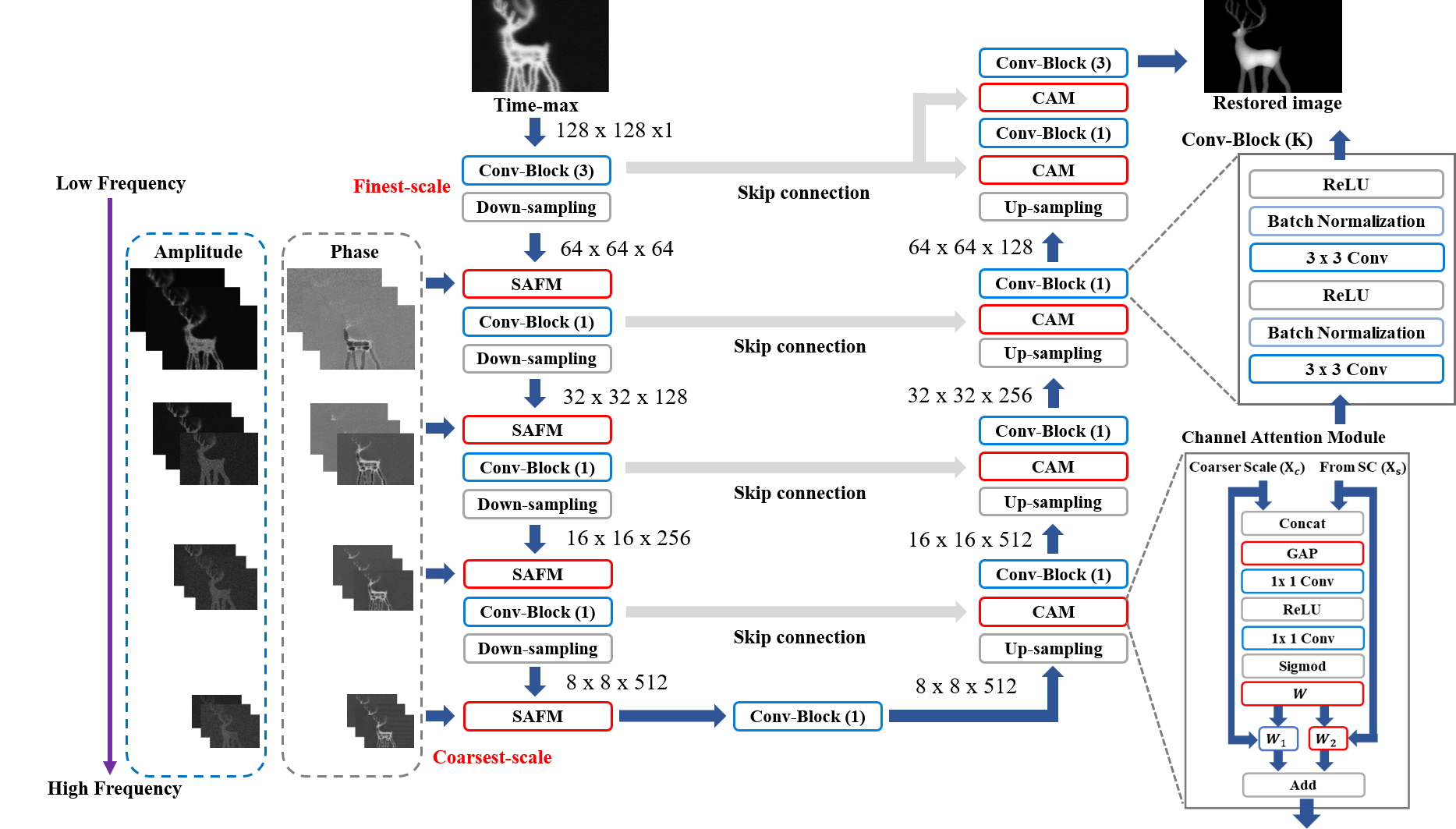}
\vspace{-0.15in}
\caption{Network architecture of \texttt{SARNet} consisting of five scale branches, where the finest-scale scale takes the \texttt{Time-max} image as input, and each of the second to fifth takes 6 images of spectral frequencies (3 bands in amplitude and 3 bands in phase) as inputs. The two gray blocks show the detailed structures of the Conv-block and the Channel Attention Module (CAM).} 
\label{fig:unet}	
\vspace{-0.2in}
\end{figure*}


\subsection{Tomographic Reconstruction}
Computer tomographic (CT) imaging methods started from X-ray imaging, and many methods of THz imaging are similar to those of X-ray imaging. One of the first works to treat X-ray CT as an image-domain learning problem was \cite{kang2017deep}, that adopts CNN to refine tomographic images. In \cite{jin2017deep}, U-Net was used to refine image restoration with significantly improved performances.  \cite{zhu2018image} further projects sinograms measured directly from X-ray into higher-dimensional space and uses domain transfer to reconstruct images. The aforementioned works were specially designed for X-ray imaging. 

Hyperspectral imaging \cite{schultz2001hyperspectral, ozdemir2020deep, geladi2004hyperspectral} constitutes  image modalities other than THz imaging. Different from THz imaging, Hyperspectral imaging collects continuous spectral band information of the target sample. Typically, the frequency bands fall in the visible and infrared spectrum; hence, most hyperspectral imaging modalities can only observe the surface characteristics of targeted objects.


\begin{figure*}[!hbt]
\centering
\includegraphics[width=0.8\textwidth]{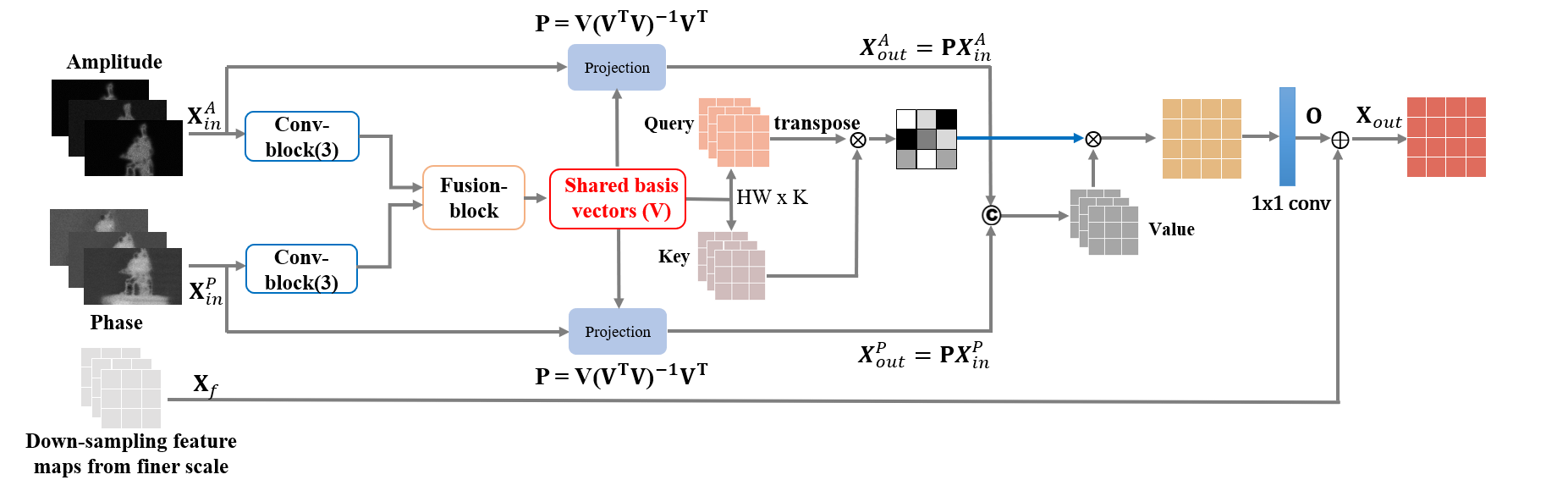}
\vspace{-0.15in}
\caption{Block diagram of Subspace-and-Attention guided Fusion Module (SAFM).} 
\label{fig:s3af}	
\vspace{-0.2in}
\end{figure*}

%% file: method.tex
\subsection{Overview}
\label{sec:overview}

As different EM bands interact with objects differently, THz waves can partially penetrate through various optically opaque materials and carry hidden material tomographic information along the traveling path. This unique feature provides a new approach to visualize the essence of 3D objects, which other imaging modalities cannot achieve. Although existing deep neural networks can learn spatio-spectral information from a considerable amount of spectral cube data, we found that directly learning from the \textbf{full spectral information} to restore THz images usually leads to an unsatisfactory performance. The main reason is that the full spectral bands of THz signals involve diverse characteristics of materials, noises, and scattered signal, which causes difficulties in model training. To address this problem, our work is based on extracting \textbf{complementary information} from both the amplitude and phase of a THz signal. That is, as illustrated in Fig.~\ref{fig:band}, in the low-frequency bands, the amplitude images delineate finer edges and object contours while the phase images offer relatively precise depths of object surfaces. In contrast, in the high-frequency bands, the amplitude images offer object mask information while the phase images delineate finer edges and object contours. Therefore, the amplitude and phase complement to each other in both the low and high frequency bands.

Motivated by the above findings, we devise a novel multi-scale Subspace-and-Attention-guided Restoration Network (\texttt{SARNet}) to capture such complementary spectral characteristics of materials to restore damaged 2D THz images effectively. The key idea of \texttt{SARNet} is to fuse spatio-spectral features with different characteristics on a common ground via deriving the shared latent subspace and discovering the short/long-range dependencies between the amplitude and phase to guide the feature fusion. To this end, \texttt{SARNet} is based on U-Net\cite{ronneberger2015u} to perform feature extraction and fusion in a multi-scale manner. 

\vspace{-0.15in}
\subsection{Network Architecture}
\label{sec:network}
On top of U-Net~\cite{ronneberger2015u}, the architecture of \texttt{SARNet} is depicted in Fig.~\ref{fig:unet}.  Specifically, \texttt{SARNet} is composed of an encoder with 5 branches of different scales (from the finest to the coarsest) and a decoder with 5 corresponding scale branches. Each scale branch of the encoder involves a Subspace-and-Attention-guided Fusion module (SAFM), a convolution block (Conv-block), and a down-sampler, except for the finest-scale branch that does not employ SAFM. To extract and fuse multi-spectral features of both amplitude and phase in a multi-scale manner,  the encoder takes a  THz 2D \texttt{Time-max} image as the input of the finest-scale branch  as well as receives to its second to fifth scale branches 24 images of additional predominant spectral frequencies extracted from the THz signal, where each branch takes 6 images of different spectral bands (3 bands of amplitude and 3 bands of phase) to extract learnable features from these spectral bands. To reduce the number of model parameters, these 24 amplitude and phase images (from low to high frequencies) are downsampled to 4 different resolutions and fed into the second to fifth scale branches in a fine-to-coarse manner as illustrated in Fig.~\ref{fig:unet}. We then fuse the multi-spectral amplitude and phase feature maps in each scale via the proposed SAFM that learns a common latent subspace shared between the amplitude and phase features to facilitate associating the short/long-range amplitude-phase dependencies. Projected into the shared latent subspace, the spectral features of amplitude and phase components, along with the down-sampled features of the upper layer, can then be properly fused together on a common ground in a fine-to-coarse fashion to obtain the final latent code.

The Conv-block($L$) contains two stacks of $L \times L$ convolution, batch normalization, and ReLU operations. Because the properties of the spectral bands of amplitude and phase can be very different, we partly use $L=1$ to learn the best linear combination of multi-spectral features to avoid noise confusion and reduce the number of model parameters. The up-sampler and down-sampler perform $2\times$ and $\frac{1}{2} \times$ scaling, respectively. The skip connections~(SC) directly pass the feature maps of different spatial scales from individual encoder branches
to the Channel Attention Modules (CAMs) of their corresponding branches of the decoder,  as indicated by the gray arrows in  Fig.~ \ref{fig:unet}. The details of SAFM and CAM will be elaborated later.

In the decoder path, each scale branch involves a up-sampler, a CAM, and a Conv-block. The Conv-block has the same functional blocks as that in the encoder. Each decoding-branch receives a ``shallower-layer''  feature map from the corresponding encoding-branch via the skip-connection shortcut and concatenates the feature map with the upsampled version of the decoded ``deeper-layer''  feature map from its coarser-scale branch.  Besides, the concatenated feature map is then processed by CAM to capture the cross-channel interaction to complement the local region for restoration. 


Note, a finer-scale branch of \texttt{SARNet} extracts shallower-layer features which tend to capture low-level features, such as colors and edges. To complement the  \texttt{Time-max} image for restoration, we feed additional amplitude and phase images of low to high spectral-bands into the fine- to coarse-scale branches of \texttt{SARNet}.  Since the spectral bands of THz amplitude and phase offer complementary information, as mentioned in Sec.~\ref{sec:overview}, besides the \texttt{Time-max} image \texttt{SARNet} also extracts multi-scale features from the amplitude and phase images of 12 selected THz spectral bands, which are then fused by the proposed SAFM.

\vspace{-0.02in}


\subsection{Subspace-and-Attention guided Fusion Module}
\label{sec:S3AF}
How to properly fuse the spectral features of THz amplitude and phase are, however, not trivial, as their characteristics are very different. To address the problem, inspired by \cite{cheng2021nbnet} and \cite{zhang2019self}, we propose the SAFM as shown in Fig.~\ref{fig:s3af}. 

We denote $\X^{A}_{\text{in}}$, $\X^{P}_{\text{in}}$ $\in$~$\mathbb{R}^{H \times W \times 3}$  as the spectral features of the THz amplitude and phase, respectively. The Conv-block $f_C(\cdot)$  extracts two intermediate feature maps $f_C(\X^{A}_{\text{in}})$, $f_C(\X^{P}_{\text{in}})$ $\in$~$\mathbb{R}^{H \times W \times C_1}$ from $\X^{A}_{\text{in}}$ and $\X^{P}_{\text{in}}$, respectively. As a result, we then derive the $\mathbf{k}$ shared basis vectors $\textbf{V}=[\textbf{v}_1, \textbf{v}_2, ..., \textbf{v}_K]$ from $f_C(\X^{A}_{\text{in}})$ and $f_C(\X^{P}_{\text{in}})$, where $\textbf{V} \in \mathbb{R}^{N \times K}$, $N=HW$ denotes the dimension of each basis vector, and $K$ is the rank of the shared subspace. The basis set of the shared common subspace is expressed as

\vspace{-0.1in}
\begin{equation}
    \textbf{V} = f_F(f_C(\X^{A}_{\text{in}}), f_C(\X^{P}_{\text{in}})),
\label{eq:basis}
\end{equation}
where we first concatenate the two feature maps in the channel dimension and then feed the concatenated feature into the fusion-block $f_F(\cdot)$. The structure of the fusion-block is the same as that of the Conv-block with $K$ output channels as indicated in the red block in Fig.~\ref{fig:s3af}. The weights of the fusion-block are learned in the end-to-end training stage. The shared latent subspace 
learning mainly serves two purposes: 1) learning the common representation between the THz amplitude and phase, and 2) learning the subspace projection matrix to project the amplitude and phase features into a shared subspace such that they can be analyzed on a common ground. These both help identify short/long-range dependencies of amplitude and phase features for feature fusion in the next stage.

To identify the short/long-range dependencies between the amplitude and phase features on a common ground, we utilize the orthogonal projection matrix $\textbf{V}$  in \eqref{eq:basis} to estimate the self-attentions in the shared feature subspace as
\begin{equation}
    \beta_{j,i} = \frac{\exp(s_{ij})}{\sum_{i=1}^{N}\exp(s_{ij})}~, ~s_{ij}= \mathbf{v}_i^{T} \mathbf{v}_j
\label{eq:correlation}
\end{equation}
where $\beta_{j,i}$ represents the model attention in the $i$-th location of the $j$-th region. The orthogonal projection matrix $\textbf{P}$ is derived from the subspace basis $\textbf{V}$ as follows \cite{meyer10matrix}:
\vspace{-0.10in}
\begin{equation}
    \textbf{P} = \textbf{V}(\textbf{V}^T\textbf{V})^{-1} \textbf{V}^T
\label{eq:orth}
\end{equation}
where $(\textbf{V}^T\textbf{V})^{-1}$ is the normalization term to make the basis vectors orthogonal to each other during the basis generation process.
As a result, the output of the self-attention mechanism becomes
\begin{equation}
    \mathbf{o}_{j} = \left( \sum_{i=1}^{N} \beta_{j,i} \mathbf{s}_{i}\right),~ ~\mathbf{s}_{i}=\texttt{Concate}(\textbf{P} \X^{A}_{\text{in}}, \textbf{P} \X^{P}_{\text{in}})
\label{eq:self}
\end{equation}
where the key of $\mathbf{s}_{i}$ $\in$~$\mathbb{R}^{HW \times 6}$ is obtained by concatenating the two feature maps  $\textbf{PX}^{A}_{\text{in}}$ and  $\textbf{PX}^{P}_{\text{in}}$ projected by orthogonal projection matrix $\textbf{P}$ $\in$~$\mathbb{R}^{HW \times HW}$, and $\X^{A}_{\text{in}}$ and $\X^{P}_{\text{in}}$ are reshaped to $HW \times 3$. Since the operations are purely linear with some proper reshaping, they are differentiable.

Finally, we further fuse cross-scale features in the self-attention output by adding the down-sampled feature map $\mathbf{X}_f$ from the finer scale as
\begin{equation}
    \mathbf{X_{out}}=f_s(\mathbf{o}) + \mathbf{X}_f
\label{eq:s3af_r}
\end{equation}
where $f_s$ is the 1×1 convolution to keep the channel number consistent with $\mathbf{X}_f$.
\subsection{Channel Attention Module}
\label{sec:CA}

To fuse multi-scale features from different spectral bands in the channel dimension, we incorporate the efficient channel attention mechanism proposed in \cite{qin2020ffa} in the decoder path of \texttt{SARNet} as shown in Fig.~\ref{fig:unet}. In each decoding-branch, the original U-Net directly concatenates the up-sampled feature from the coarser scale with the skip-connection feature from the corresponding encoding-branch, and then  fuse the intermediate features from different layers by convolutions. This, however, leads to poor image restoration performances in local regions such as incorrect object thickness or details. To address this problem, we propose a channel attention module (CAM) which adopts full channel attention in the dimensionality reduction operation by concatenating two channel attention groups. CAM  first performs the global average pooling to extract the global spatial information in each channel:
\vspace{-0.1in}
\begin{align}
    G_t = \frac{1}{H \times W}\sum_{i=1}^{H}\sum_{j=1}^{W} X_{t}(i,j)
\label{eq:avg}
\end{align}
where $X_{t}(i,j)$ is the $t$-th channel of $X_{t}$ at position $(i,j)$ obtained by concatenating the up-sampled feature map $\textbf{X}_c$ of the coarser-scale and the skip-connection feature map $\textbf{X}_s$. The shape of $G$ is from $C \times H \times W$ to  $C \times 1 \times 1$. 


We directly feed the result through two $1\times 1$ convolution, sigmoid, and ReLU activation function as:

\vspace{-0.2in}
\begin{align}
    \textbf{w} = \sigma \left( \texttt{Conv} \left( \delta \left( \texttt{Conv}(G) \right) \right) \right),
\label{eq:channel_comp}
\end{align}
where $\texttt{Conv}$ represents the $1\times 1$  convolution, $\sigma$ is the sigmoid function, and $\delta$ is the ReLU function.
In order to better restore a local region, we group the weights $\textbf{w}$ of different channels as two groups $\textbf{w}=[\textbf{w}_1, \textbf{w}_2]$ corresponding to two different sets of input feature maps. Finally, we element-wise multiply the input $X_c$, $X_s$ of the weights $\textbf{w}$ and add these two groups features.

\subsection{Loss Function for THz Image Restoration}
\label{sec:loss}
To effectively train \texttt{SARNet}, we employ the following mean squared error (MSE) loss function to measure the dissimilarity between the restored image $\mathbf{X}_{\text{rec}}$ and its ground-truth $\mathbf{X}_{\text{GT}}$:
	\begin{align}
	\mathcal{L}_{\text{MSE}}(\mathbf{X}_{\text{GT}},\mathbf{X}_{\text{rec}}) = &\frac{1}{HW}\sum_{i=1}^{H}\sum_{j=1}^{W}(\mathbf{X}_{\text{GT}}(i,j)-\mathbf{X}_{\text{rec}}(i,j))^2,
	\label{eq:loss}
	\end{align}
where  $H$ and $W$ are the height and width of the image. 

The 3D tomography of an object can then be reconstructed from the restored THz 2D images of the object scanned in different angles. To this end, we can directly apply the inverse Radon transform to obtain the 3D tomography, using methods like filtered back-projection \cite{kak2001algorithms} or the simultaneous algebraic reconstruction technique \cite{recur2011investigation}.

%% file: experiments.tex
\begin{figure}[t]
\centering
\includegraphics[width=0.45\textwidth]{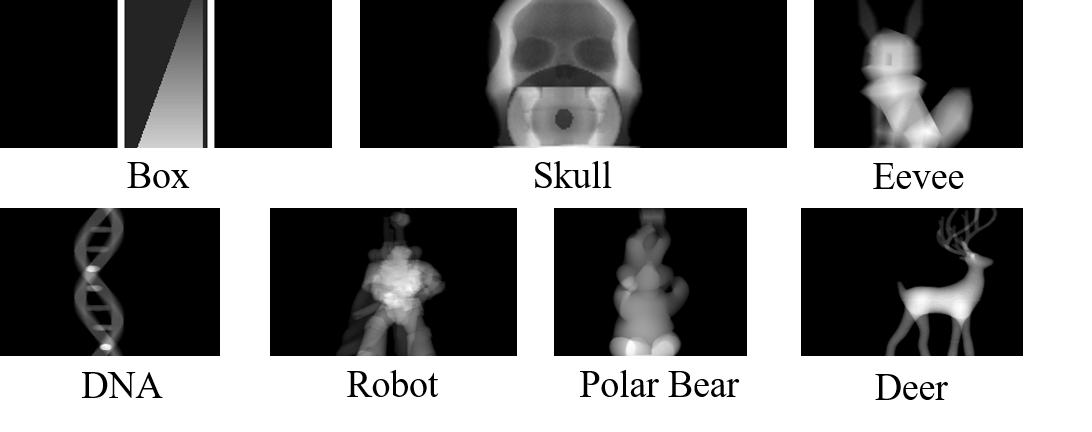}
\vspace{-0.2in}
\caption{Illustration of authentic measured THz data for the seven 3D-printed HIPS objects in our experiments.} 
\label{fig:data}	
\vspace{-0.2in}
\end{figure}

\begin{figure*}[!hbt]
\centering
\includegraphics[width=1\textwidth]{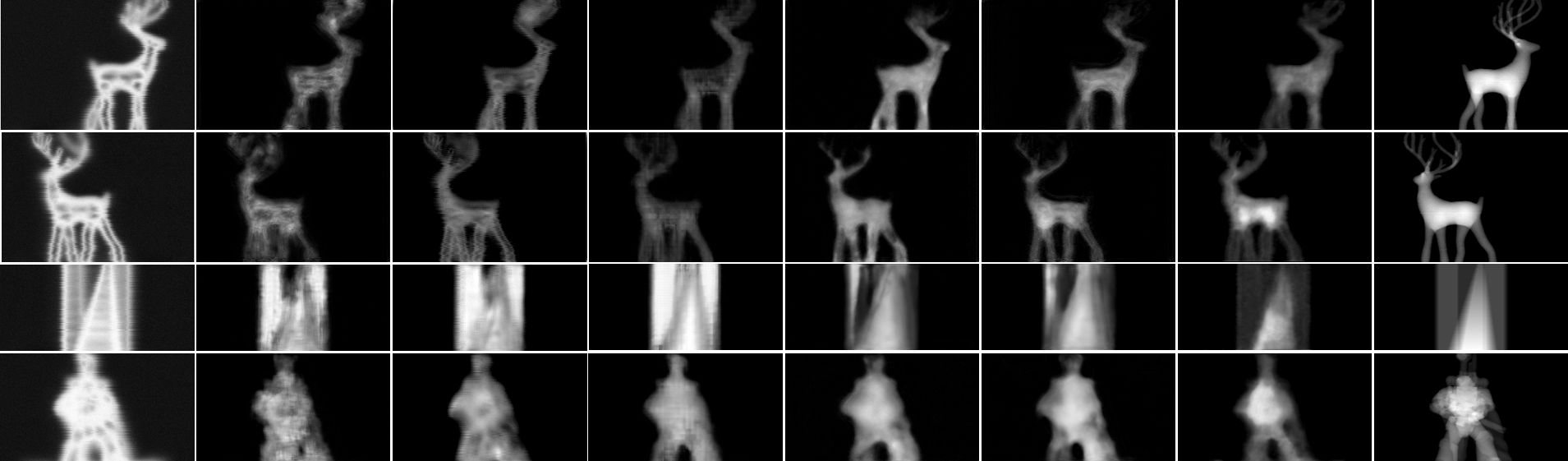}
\vspace{-0.3in}
\caption{Qualitative comparison of THz image restoration results for \textbf{Deer}, \textbf{Box}, and  \textbf{Robot} from left to right: (a) $\texttt{Time-max}$, (b) $\texttt{DnCNN-S}$ \cite{zhang2017beyond}, (c) $\texttt{RED}$ \cite{mao2016image}, (d) $\texttt{NBNet}$ \cite{cheng2021nbnet}, (e) $\texttt{U-Net}_\texttt{base}$ \cite{ronneberger2015u}, (f) $\texttt{U-Net}_\texttt{MS}$, (g) $\texttt{SARNet}$,  and (h) the ground-truth.} 
\label{fig:vis1}	
\vspace{-0.25in}
\end{figure*}

We conduct experiments to evaluate the effectiveness of \texttt{SARNet} against existing state-of-the-art restoration methods. We first present our experiment settings and then evaluate the performances of \texttt{SARNet} and the competing methods on THz image restoration.

\subsection{THz-TDS Image Dataset}
As shown in Fig.~\ref{fig:design}, we prepare the  sample objects by a Printech 3D printer, and use the material of high impact polystyrene (HIPS) for 3D-printing the objects due to its high penetration of THz waves. We then use our in-house Asynchronous Optical Sampling THz time-domain spectroscopy (ASOPS THz-TDS) system~\cite{janke2005asynchronous} to measure the sample objects. Each sample object is placed on a motorized stage between the source and the receiver. With the help of the motorized stage, raster scans are performed on each object in multiple view angles. In the scanning phase, we scan the objects covering a rotational range of 180 degrees (step-size: 6 degrees),  a horizontal range of 72mm (step-size: 0.25mm), and a variable vertical range corresponding to the object height (step-size: 0.25mm). In this way, we obtain 30 projections of each object, which are then augmented to 60 projections by horizontal flipping. The ground-truths of individual projections are obtained by converting the original 3D printing files into image projections in every view-angle. We use markers to indicate the center of rotation so that we can align the ground-truths with the measured THz data. In this paper, totally 7 objects are printed, measured, and aligned for evaluation. 

\vspace{-0.05in}
\subsection{Data Processing and Augmentation}
In our experiments, we train  \texttt{SARNet} using the 2D THz images collected from our THz imaging system shown in Fig.~\ref{fig:band}. The seven sample objects are illustrated in Fig.~\ref{fig:data}, consisting of 60 projections per object and 420 2D THz images in total.  In order to thoroughly evaluate the capacity of \texttt{SARNet}, we adopt the leave-one-out strategy: using the data of 7 objects as the training set, and that of the remaining object as the testing set. Due to the limited space, we only present part of the results in this section, and the complete results in the supplementary material. We will release our code and the THz image dataset after the work is accepted.

We also perform typical data augmentations to enrich the training set, involving random transformations on brightness, contrast, and saturation,image scaling with a scaling factor ranging in $[0.5, 1.5]$, and geometric transformations including random flipping horizontally and vertically. Finally, the images are randomly cropped to $128 \times 128$ patches.

\subsubsection{Experiment Settings}
We initialize \texttt{SARNet} following the initialization method in \cite{he2015delving}, and train it using the Adam optimizer with $\beta_1 = 0.9$ and $\beta_2 = 0.999$. We set the initial learning rate to $10^{-4}$ and then decay the learning rate by $0.1$ every $300$ epochs. \texttt{SARNet} converges after $1,000$ epochs.  For a fair comparison with the competing methods, we adopt their publicly released codes.

\subsection{Quantitative Evaluations}
\vspace{-0.1in}

To the best of our knowledge, there is no method  specially designed for restoring THz images besides \texttt{Time-max}. Thus, we compare our method against several representative CNN-based image restoration models, including DnCNN~\cite{zhang2017beyond}, RED \cite{mao2016image}, and NBNet~\cite{cheng2021nbnet}. Moreover, we also compare two variants of U-Net \cite{ronneberger2015u}: baseline U-Net ($\texttt{U-Net}_\texttt{base}$) and multi-spectral U-Net ($\texttt{U-Net}_\texttt{MS}$). $\texttt{U-Net}_\texttt{base}$  extracts image features in five different scales following the original setting in U-Net \cite{ronneberger2015u}, whereas $\texttt{U-Net}_\texttt{MS}$ incorporates multi-spectral features by concatenating the features of \texttt{Time-max} image with additional 12 THz bands for amplitude as the input (i.e., $12+1$ channels) of the finest scale of U-Net. 
For objective quality assessment, we adopt two widely-used metrics including the Peak Signal-to-Noise Ratio (PSNR) and Structural SIMilarity (SSIM) to respectively measure the pixel-level and structure-level similarities between a restored image and its ground-truth.

Table\,\ref{tab:t1} shows that our \texttt{SARNet} significantly outperforms the competing methods on all sample objects in both metrics. 
Specifically, \texttt{SARNet} outperforms $\texttt{Time-max}$,  baseline U-Net ($\texttt{U-Net}_\texttt{base}$), and the multi-spectral U-Net ($\texttt{U-Net}_\texttt{MS}$) by 11.17 dB, 2.86 dB, and 1.51 dB in  average PSNR, and 0.744, 0.170, and 0.072 in  average SSIM for 7 objects. For qualitative evaluation, Fig.\,\ref{fig:vis1} illustrates a few restored views for \textbf{Deer}, \textbf{Box}, and \textbf{Robot}, demonstrating that \texttt{SARNet} can restore objects with much finer and smoother details (e.g., the antler and legs of \textbf{Deer}, the depth and shape of \textbf{Box}, and the body of \textbf{Robot}),  faithful thickness of material (e.g., the body and legs of \textbf{Deer} and the correct edge thickness of \textbf{Box}), and fewer artifacts (e.g., holes and broken parts). Both the quantitative and qualitative evaluations confirm a significant performance leap with \texttt{SARNet} over the competing methods.


	\begin{table}[!t]
		\caption{Quantitative comparison (PSNR and SSIM) of THz image restoration performances with different methods on \textbf{Deer}, \textbf{DNA}, \textbf{Box}, \textbf{Eevee}, \textbf{Polarbear}, \textbf{Robot}, and \textbf{Skull}.  ($\uparrow$: higher is better)}
		\begin{center}
		    \vspace{-0.2in}
			\begin{scriptsize}
			\scriptsize
			\scalebox{0.82}{
				\begin{tabular}{|c|c|c|c|c|c|c|c|} \hline
				    Method & \multicolumn{7}{c|}{PSNR$\uparrow$} \\
				    \cline{2-8}
					                                     &Deer   &DNA    &Box     &Eevee    &Polarbear    &Robot   &Skull  \\ \hline
				    $\texttt{Time-max}$                             &12.42  &12.07  &11.97   &11.20    &11.21        &11.37   &10.69 \\ \hline
					$\texttt{DnCNN-S}$ \cite{zhang2017beyond}       &19.94  &23.95  &19.13   &19.69    &19.44        &19.72   &17.33  \\ \hline
					$\texttt{RED}$ \cite{mao2016image}              &19.30	 &24.17  &20.18	  &19.97    &19.17        &19.76   &16.28  \\ \hline
					$\texttt{NBNet}$ \cite{cheng2021nbnet}          &20.24  &25.10  &20.21   &19.84    &20.12        &20.01      &19.69     \\ \hline
					$\texttt{U-Net}_\texttt{base}$ \cite{ronneberger2015u}    &19.84	 &24.15  &19.77	  &19.95    &19.09        &18.80   &17.49  \\ \hline
					$\texttt{U-Net}_\texttt{MS}$                              &22.46	 &25.05  &20.81	  &20.34    &19.86        &20.64   &19.43  \\ \hline 
					$\texttt{SARNet}$ (Ours)                     &\textbf{22.98}	 &\textbf{26.05}  &\textbf{22.67}	 &\textbf{20.87}  &\textbf{21.42}   &\textbf{22.66}   &\textbf{22.48}   \\ \hline \hline
                     Method  & \multicolumn{7}{c|}{SSIM$\uparrow$} \\
                     \cline{2-8}
					                                     &Deer   &DNA    &Box     &Eevee    &Polarbear    &Robot   &Skull  \\ \hline
				    $\texttt{Time-max}$                             &0.05   &0.05   &0.14  &0.14     &0.12        &0.08    &0.09   \\ \hline
					$\texttt{DnCNN-S}$ \cite{zhang2017beyond}       &0.73   &0.77   &0.73  &0.72     &0.63        &0.77    &0.36   \\ \hline
					$\texttt{RED}$ \cite{mao2016image}              &0.81   &0.83   &0.74  &0.77     &0.75        &0.80    &0.74   \\ \hline
					$\texttt{NBNet}$ \cite{cheng2021nbnet}          &0.81   &0.85   &0.75  &0.77     &0.43        &0.80      &0.78     \\ \hline
					$\texttt{U-Net}_\texttt{base}$ \cite{ronneberger2015u}    &0.55   &0.78   &0.77  &0.76     &0.56        &0.76    &0.51   \\ \hline
					$\texttt{U-Net}_\texttt{MS}$                              &0.76   &0.73   &0.78  &0.76     &0.78        &0.79    &0.78   \\ \hline 
					$\texttt{SARNet}$ (Ours)                      &\textbf{0.84}  &\textbf{0.90} &\textbf{0.83}   &\textbf{0.82}   &\textbf{0.82}   &\textbf{0.83}   &\textbf{0.84}  \\ \hline
				\end{tabular}}
			\end{scriptsize}
		\end{center}
		\label{tab:t1}
 		\vspace{-0.25in}
	\end{table}

		\begin{table}
		\caption{Quantitative comparison (PSNR and SSIM) of THz image restoration performances on \textbf{Deer}, \textbf{DNA} and \textbf{Box} under different settings. ($\uparrow$: higher is better)}
		\begin{center}
		    \vspace{-0.20in}
			\begin{scriptsize}
			\scriptsize
			\scalebox{0.9}{
				\begin{tabular}{|c|c|c|c|c|c|c|} \hline
				    Method & \multicolumn{3}{c|}{PSNR$\uparrow$} & \multicolumn{3}{c|}{SSIM$\uparrow$} \\
				    \cline{2-4} \cline{5-7} 
					                                               &Deer     &DNA    &Box   &Deer   &DNA   &Box   \\ \hline\hline
					$\texttt{U-Net}_\texttt{base}$ \cite{ronneberger2015u}  &19.84	 &25.63  &19.77 &0.55   &0.78  &0.77\\ \hline
					Amp-Unet w/o attention                         &22.05	 &25.84  &20.32 &0.80   &0.83  &0.77   \\ \hline
                    Phase-Unet w/o attention                       &21.14	 &24.98  &20.42 &0.82   &0.72  &0.78  \\ \hline
                    Mix-Unet w/o attention                         &21.44    &25.78  &20.00 &0.81   &0.81  &0.78  \\ \hline
                    Amp-Unet w/ attention                          &20.97    &26.00  &21.83 &0.84   &0.90  &0.78  \\ \hline
                    Phase-Unet w/ attention                        &22.66    &25.52  &21.65 &0.83   &0.86  &0.79  \\ \hline
					$\texttt{SARNet}$ (Ours)                   &\textbf{22.98}	 &\textbf{26.05}  &\textbf{22.67}  &\textbf{0.84}  &\textbf{0.90}   &\textbf{0.83} \\ \hline
				\end{tabular}}
			\end{scriptsize}
		\end{center}
		\label{tab:t2}
 		\vspace{-0.4in}
	\end{table}
	

\subsection{Ablation Studies}
To verify the  effectiveness of multi-spectral feature fusion, we evaluate the restoration performances with our \texttt{SARNet} under different settings  in Table\,\ref{tab:t2}. The compared methods include (1) $\texttt{U-Net}_\texttt{base}$ using a single channel of data (\texttt{Time-max}) without using features of multi-spectral bands; (2) \textbf{Amp-Unet w/o attention} employing multi-band amplitude feature (without the attention mechanism) in each of the four spatial-scale branches, except for the finest scale  (that accepts the \texttt{Time-max} image as the input), where  12 spectral bands of amplitude (3 bands/scale) are fed into the four spatial-scale branches with the assignment of the highest-frequency band to the coarsest scale, and vice versa; (3) \textbf{Phase-Unet w/o attention} employing multi-spectral phase features with the same spectral arrangements as (2), and without the attention mechanism; (4) \textbf{Mix-Unet w/o attention} concatenating multi-spectral amplitude and phase features (without the attention mechanism) in each of the four spatial-scale branches, except for the finest scale (that accepts the Time-max image as the input), where  totally 24 additional spectral bands of amplitude and phase (3 amplitude plus 3 phase bands for each scale) are fed into the four branches; (5) \textbf{Amp-Unet with attention} utilizing attention-guided  multi-spectral amplitude features with  the same spectral arrangements as specified in (2); and (6) \textbf{Phase-Unet with attention} utilizing attention-guided multi-spectral phase features with the same spectral arrangements as in (2).

The results clearly demonstrate that the proposed SAFM can benefit fusing  the spectral features of both amplitude and phase with different characteristics for THz image restoration. 
Specifically, employing additional multi-spectral features of either amplitude or phase as the input of the multi-scale branches in the network (i.e., Amp-Unet or Phase-Unet w/o attention) can achieve performance improvement over $\texttt{U-Net}_\texttt{base}$.  Combining both the amplitude and phase features without the proposed subspace-and-attention guided fusion (i.e., \textbf{Mix-Unet w/o attention}) does not outperform \textbf{Amp-Unet w/o attention} and usually leads to worse performances. The main reason is that the characteristics of the amplitude and phase features are too different to be fused to extract useful features with direct fusion methods. This motivates our subspace-and-attention guided fusion scheme, that learns to effectively identify and fuse important and complementary features on a common ground. 

\vspace{-0.1in}
\begin{figure}[t]
\centering
\includegraphics[width=0.50\textwidth]{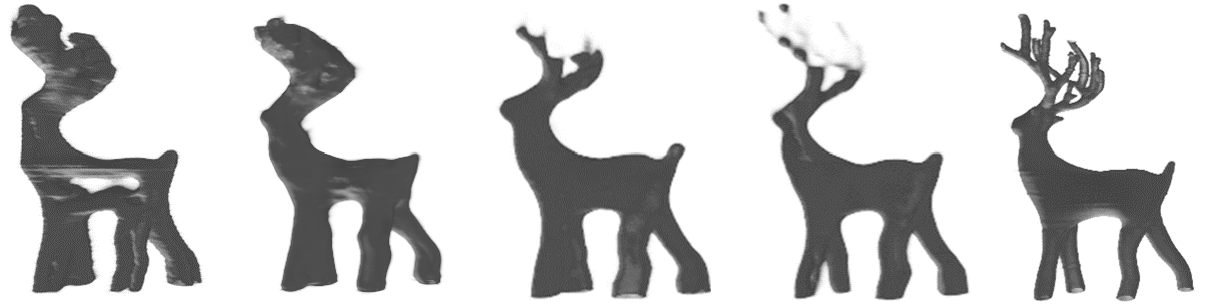}
\vspace{-0.3in}
\caption{Illustration of 3D tomographic reconstruction results on \textbf{Deer},  from left to right: (a) $\texttt{Time-max}$, (b) $\texttt{U-Net}_\texttt{base}$~\cite{ronneberger2015u}, (e) $\texttt{U-Net}_\texttt{MS}$, (d) $\texttt{SARNet}$, and (e) ground-truth.} 
\label{fig:3d}	
\vspace{-0.25in}
\end{figure}

    

\subsection{3D Tomography Reconstruction}
Our goal   is  to  reconstruct clear and faithful  3D  object shapes  through  our THz tomographic  imaging  system. In our system, the tomography of an object is reconstructed from 60 views of 2D THz images of the object, each being restored by \texttt{SARNet}, via the inverse Radon transform. Fig.~\ref{fig:3d} illustrates the 3D reconstructions of \textbf{Deer}, showing that \texttt{Time-max},  $\texttt{U-Net}_\texttt{base}$, and $\texttt{U-Net}_\texttt{MS}$ tend to lose important object details such as holes in the deer's body with \texttt{Time-max} and the severely distorted antlers and legs with the three methods. In contrast, our method reconstructs much clearer and more faithful 3D images with finer details, achieving by far the best 3D THz tomography reconstruction quality in the literature. Complete 3D reconstruction results are provided in the supplementary material.



%% file: conclusion.tex
We proposed a 3D THz imaging system based on fusing multi-spectral features which is the first to merge THz spatio-spectral data, data-driven models, and material properties. Based on the physical characteristics of THz waves passing through different materials, our novel \texttt{SARNet} efficiently fuses spatio-spectral features with different characteristics on a common ground via deriving a shared latent subspace and discovering the short/long-range dependencies between the amplitude and phase to guide the feature fusion for boosting restoration performance. Our results have confirmed a performance leap from the relevant state-of-the-art techniques in the area.  We believe our findings in this work will stimulate further applicable research for THz imaging with advanced computer vision techniques.